\documentclass[twocolumn,a4paper,reprint,amssymb,aps,prd,groupedaddress,superscriptaddress,amsmath,nobibnotes,showpacs,nofootinbib]{revtex4-2}
\pdfoutput=1
\usepackage{graphicx,bm,color,psfrag}
 \usepackage{bigints} 
\usepackage{amsfonts}
\usepackage{lipsum}
\usepackage{enumitem}
\usepackage[colorlinks = true,
            linkcolor = blue,
            urlcolor  = blue,
            citecolor = blue,
            anchorcolor = blue]{hyperref}

\usepackage{appendix}
\usepackage{xcolor}
\usepackage{threeparttable,booktabs}

\begin{document}

\title{
Role of internal space correlations in the dynamics of a higher-dimensional \\ Bianchi type-I universe: shear scalar and Hubble parameter perspectives}

\author{Nihan Kat{\i}rc{\i}}
\email{nkatirci@dogus.edu.tr}
\affiliation{Department of Electrical-Electronics Engineering Do\u gu\c s University, \" Umraniye 34775 Istanbul, Türkiye}

\begin{abstract}

We investigate exact solutions of the Einstein field equations in higher-dimensional, spatially homogeneous Bianchi type-I spacetimes, introducing a real parameter $\lambda$ that correlates the expansion rates of external and internal spaces. Extending beyond Robertson–Walker spacetime, our approach includes positive and negative correlations, suggesting a broader and isotropic/anisotropic cosmological model space. Positively correlated dimensions manifest as a cosmological constant at late times, while at early times, they mimic stiff-fluid-like dark energy that dilutes faster than radiation, paralleling early dark energy models. This suggests a pathway for alleviating the Hubble tension by tailoring higher-dimensional dynamics to reduce the sound horizon. When anisotropic expansion is allowed, these models achieve isotropization more efficiently than predicted by Wald’s cosmic no-hair theorem. Negative correlations, in contrast, yield a higher-dimensional steady-state universe where the shear scalar remains constant, effectively emulating a negative cosmological constant. These distinct behaviors arise from a simple signature change: positive correlation accelerates shear scalar decay, while negative correlation stabilizes it. We demonstrate that the solutions admit analytic continuation from the Lorentzian to Euclidean regime ($t \rightarrow -i\tau$), revealing a wormhole-like topology that connects two asymptotic regions via a throat, with $\lambda \rightarrow -\lambda$.
\end{abstract}

\maketitle

\section{Introduction}
\label{sec:intro}
Compact/non-compact higher dimensions have been extensively used in addressing the problems in cosmology [e.g., inflation, dark matter and dark energy] and in particle physics [e.g., neutrino masses, proton instability], even connecting different large hierarchies appearing in both, expected to be described by a complete fundamental theory of physics. Contrariwise, motivation can be totally theoretical such that string \cite{Goto:1971ce,Green:1987sp} or string inspired theories, e.g., ghost-free Gauss-Bonnet gravity require higher dimensions \cite{Zwiebach:1985uq,Nojiri:2006je}, their predictions can be utilized in the sense of addressing these problems as well.

Now, we employ a higher dimensional Bianchi type-I metric with the assumption that the internal space is homogeneous and isotropic. Then we present solutions with a \textit{constant} correlation between the external (observable) and internal spaces tracing possible changes on the expansion dynamics $H(z)$ and its expansion anisotropy $\sigma_{\rm ext}^2(z)$. The sequel paper of \cite{akarsuisotropic,Akarsu:2014dxa} (published in $2013$ and $2015$) in which we extend a class of isotropic cosmological solutions of higher dimensional Einstein field equations with the source having the energy-momentum tensor (EMT) of a homogeneous, isotropic fluid by replacing a spatially flat, isotropic and homogeneous Robertson-Walker (RW) metric for external space with a homogeneous and anisotropic Bianchi type-I metric is required in the literature. Since in \cite{akarsuisotropic}, the authors postulate a kinematical constraint to correlate the expansion rates of the external and internal spaces in terms of a real parameter $\lambda$, staying loyal to \textit{ a positive constant} to close the system of field equations, we will show here particularly different expansion scenarios considering the current developments and cosmological tensions as well. Ref. \cite{akarsuisotropic} exhibits $\Lambda$CDM-like cosmology with expanding internal space as well, yet staying at Planck length scales even today in the presence of three-dimensional internal space $n=3$, Ref. \cite{Akarsu:2014dxa} extends the solution to a general solution for arbitrary values of $n$. We have seen that for the RW extension, for positive $\lambda$, four dimensional effective fluid becomes a conventional cosmological constant at late times, and acts like a stiff-fluid at early times which dilutes faster than radiation, in this sense it may mimic early dark energy models, then tailoring higher dimensional theories may be promising in alleviating the Hubble tension. We extend the analysis of \cite{akarsuisotropic} by incorporating negative correlation between internal and external spaces to show how significantly it deviate from its standard expansion anisotropy and its potential to yield very different cosmologies still staying in the standard GR. Surprisingly, the positive correlation leads to a drastically different isotropization. A scheme of the study is given in Fig.~\ref{fig:Scheme} in Appendix \ref{app:scheme} to assist readers in navigating this paper. 

Within four-dimensional general relativity (GR), expansion anisotropy---quantified by shear scalar ($\sigma^2$)---resembles the kinetic energy contribution of a canonical scalar field having an EoS parameter $w=1$---causality limit~\cite{EllisRC} for any physical source---thereby expansion anisotropy is the fastest diluting component with almost $\rho_{\sigma^2}\propto a^{-6}$ compared to any material sources having $w<1$ as the Universe expands, therefore we are not expecting to see the effects of anisotropy on the late universe atop of the $\Lambda$CDM (Lambda-cold dark matter) model. Generalized Friedmann equation containing the average Hubble parameter along with the shear scalar term  retaining isotropic spatial curvature has been studied ~\cite{Ellis69,Collins:1972tf,Ellis99,EllisRC,Akarsu:2019pwn,Akarsu:2021max} and is well constrained~\cite{Akarsu:2019pwn,Akarsu:2021max} in which a very small amount of present-day expansion anisotropy can not be excluded such that the data favors the isotropic expansion, with $\Omega_{\sigma0}\leq10^{-18}$ at $95\%$ confidence level from cosmic microwave background (CMB) data with lensing. As stated in \cite{Gron:2024vmf}, the anisotropic generalization of the $\Lambda$CDM model, under the stringent constraints, is not sufficient to solve the Hubble tension as well. However, an observational deviation from its stiff-fluid character in GR may imply a necessity for replacing GR by modified theories of gravity.\footnote{Modified gravity theories can be described as an imperfect fluid exerting different pressures in three directions (corresponding to different shear scalar evolutions) in this approach where the field equations are considered as effective Einstein equations, may shed light on dark energy,~\cite{Pimentel89,Madsen88,Faraoni:2018qdr,Akarsu:2019pvi,Mimoso:1995ge}.}. Similarly, the presence of extra dimensions in GR and in its alternatives requiring higher dimensions provide a fluid of geometric DE whose density parameter is determined by the number of extra dimensions, expansion anisotropy, if exists and can be detected, provides an alternative way of probing DE. For instance, the discovery of the expansion anisotropy in a very particular form different than GR predictions would be a strong signal in favor of modified theories and/or higher dimensions rather than the presence of a DE (such as $\Lambda$ and scalar fields) ingredient of the Universe assuming that GR is valid, see ~\cite{Akarsu:2020pka}. For instance, in ~\cite{Akarsu:2012am}, the higher dimensional Robertson-Walker cosmologies has been studied and then in a similar manner, its anisotropic extension \cite{Akarsu:2021fie} reveals that the shear scalar of the external space mimics a negative cosmological constant instead of stiff fluid-like evolution in the standard $\Lambda$CDM model under a higher dimensional steady-state universe ansatz. Interestingly, a higher dimensional steady-state universe gives rise to mathematically exactly the same Friedmann equation of the standard $\Lambda$CDM model for the external space, in which the higher-dimensional negative cosmological constant plays the role of the four-dimensional positive cosmological constant. In this study, we show that by allowing a negative correlation between the spaces different than \cite{akarsuisotropic}, we are able to reproduce a higher dimensional steady-state universe via the kinematical constraint as well. For instance, in~\cite{Akarsu:2020pka} it was shown that if one stays loyal to zero inertial mass density $p_{\Lambda}+\rho_{\Lambda}=0$ for DE on average allowing expansion anisotropy, shear scalar tracks the DE deforming it meaning that all cosmologies based on the canonical scalar fields should be reconsidered from this point of view, on top of that, along with a bonus, the kinetic term of the scalar field is replaced by an real observable, the shear scalar.

Besides theoretical motivations, in principle relaxing RW metric could resolve $\Lambda$CDM tensions, that means that the whole conventional model of cosmology with tensions need to be completely reconstructed. The CMB anomalies (when $\Lambda$CDM is considered) have been observed at large angular scale in the CMB maps by the COBE, WMAP and Planck satellites such as 
a hemispherical power asymmetry, an unexpected large cold spot in the Southern hemisphere, a lack of correlation on the largest angular scales, a lack of quadrupole power, a preference for odd parity modes, an alignment between various low multipole moments, an alignment between those low multipole moments and the motion and geometry of the Solar System, see~\cite{Bennett11,Ade:2013kta,Schwarz:2015cma,Akrami:2019bkn,Wilczynska:2020rxx,Migkas:2020fza} and references therein for these features in CMB as well as in other types of cosmological data~\cite{Dhawan:2022lze,Cowell:2022ehf,McConville:2023xav}. In a recent work \cite{Jones:2023ncn}, authors also argue that the anomalies listed above jointly constitute more than $5\sigma$ violation for statistical isotropy and these may be an evidence suggesting that the Universe might not be as isotropic as thought. These observations generate considerable interest, enabling a bidirectional approach where observations not only lead to theoretical formulation but also benefit from theoretical guidance. For instance, as shown in \cite{Gallego:2024gay}, anisotropic dark energy can be constructed from higher-dimensional superalgebras and superstring theories using conventional methods, or as in \cite{Verma:2024lex}, testing observational deviations in the directional equation of state parameters through specific parameterizations is also possible irrespective of the underlying theoretical motivations.

Even these findings are interpreted by the community as a possibility to address Hubble ($H_0$) tension ~\cite{DiValentino:2020zio,DiValentino:2021izs} for a recent review on the $H_0$ tension) by reanalyzing the cosmological data by breaking down of the RW framework, e.g., allowing anisotropic expansion in the late universe~\cite{Colin:2017juj,Colin:2019opb,Secrest:2020has,Krishnan:2021dyb,Krishnan:2021jmh,Luongo:2021nqh,Aluri:2022hzs}. If the Hubble parameter varies depending on the direction, previous studies assuming an isotropic Hubble expansion, excluding the direction dependence, could be biased in their results. Measurements of the shear scalar with upcoming surveys, such as the Euclid satellite~\cite{Amendola:2016saw}, or Zwicky Transition Facility SNe Ia sample probes, the first X-ray survey eROSITA could reveal more on the nature of DE. From cosmology-independent methods like cluster X-ray temperature, they detect $9\%$ anisotropy in $H_0$ in dipol directions or $900$ km/s bulk flow  \cite{Migkas:2021zdo} with statistical significance of the anisotropy to be $> 5\sigma$ and first eROSITA results independently support previously detected anisotropy in local Universe. Ultimately, galaxy clusters can say whether $H_0$ is the same in all directions and bulk flows are consistent with $\Lambda$CDM or not. 

Beyond providing different solutions, there is another aspect we discuss in the current study. We realize that in Einstein equations, 3-dimensional or higher-dimensional, Hubble parameter and its derivatives remain invariant for flat space-like sections if a space coordinate turns into a time coordinate ($a^2\rightarrow -a^2$). Since the non-ratio form of spatial curvature and spatial anisotropy breaks this invariance, the change of signature can be traced only by $\pm 1/a^2$ term---provided that the Universe is spatially curved--- or $1/a^6$-like term---provided that expansion is anisotropic in both GR and its alternatives. 

This point has been clarified from signature of $\lambda$ constant kinematical constraint. Furthermore, we also show that the solutions given in this study, as well as isotropic ones in~\cite{akarsuisotropic}, for $\lambda>0$ remain invariant under transformations $\lambda\rightarrow -\lambda$ and $t\rightarrow -i\tau$, it means that solutions with $\lambda<0$ in Euclidean metric is equivalent to solutions with $\lambda>0$ in Lorentzian metric accompanied by a signature change of correlation constant. Since correlation constant acts like a cosmological constant and its sign switch  aligns with the findings from the sign-switching $\Lambda$ model ($\Lambda_{\rm s}$CDM)~\cite{Akarsu:2021fol,Akarsu:2022typ,Akarsu:2023mfb}. Therefore this wormhole-type continuation may be used to generate a plausible mechanism for transition of the Universe from anti-de Sitter (AdS) vacua to de Sitter (dS) vacua in the late Universe at redshift $z_\dagger \sim 2$, proposed in the $\Lambda_{\rm s}$CDM model to address major cosmological tensions such as the $H_0$, $S_8$, and $M_B$ tensions simultaneously, while also accommodating the BAO Lyman-$\alpha$ discrepancy \cite{Akarsu:2019hmw}. Studies \cite{Anchordoqui:2023etp,Anchordoqui:2023woo} have shown that the $\Lambda_{\rm s}$CDM model can be related to the dark dimension scenario \cite{Montero:2022prj}, where the dependence between the vacuum energy and the radius at its minimum remains the same as before the boost. Consequently, our observable (external) space, with different correlation functions [even varying kinematical constraints, i.e., $H_{\rm ext} H_{\rm int} = f(z)$] for the internal space (whether dark or not), elucidates the capability of higher-dimensional models to alleviate tensions with different predictions (such as massive gauge bosons/fermions and neutrinos furnished by bulk masses), which serve as potential smoking guns.

\section{The $(1+3+n)$ dimensional spacetime}
We consider a minimal extension of the $(1+3)$-dimensional Einstein's general theory of gravity to $(1+3+n)$ dimensions, preserving its mathematical structure\footnote{Note that the quantities with tilde stand for the $(1+3+n)$ dimensional values following the same notation in \cite{Akarsu:2021fie} whereas the inverse notation was used in \cite{akarsuisotropic}.}, i.e., we assume that test particles follow geodesics of a $(1+3+n)$ dimensional spacetime whose geometry relates to the energy-momentum distribution on the manifold through
\begin{equation}
\begin{aligned}
\label{eqn:EFE}
\tilde{R}_{\mu\nu}-\frac{1}{2}\tilde{R}\tilde{g}_{\mu\nu} =-\tilde{\kappa}\tilde{T}_{\mu\nu},
\end{aligned}
\end{equation}
where the indices $\mu,\nu=0,1,...,3+n$ and $0$-index denotes the cosmic (proper) time, $t$. The $ \tilde{\kappa}=8\pi \tilde{G}$ where $\tilde{G}$ is the gravitational constant for $(1+3+n)$-dimensional spacetime. In Eq.\eqref{eqn:EFE}, $\tilde{g}_{\mu\nu}$ and $\tilde{R}_{\mu\nu}$ are respectively the metric tensor and the Ricci tensor, constructing $\tilde{R}=\tilde{g}^{\mu\nu}\tilde{R}_{\mu\nu}$, the Ricci curvature scalar of the $(1+3+n)$-dimensional spacetime. The spacetime manifold with product topology is
\begin{equation}
\mathcal{M}^{1+3+n}=\mathcal{R}\times \mathcal{M}^{3}\times \mathcal{K}^{n},
\end{equation}
where $\mathcal{R}$ and $\mathcal{M}^3$ are the manifold of time, and of $3$-dimensional external space that represents the space we observe,  respectively. $\mathcal{K}^{n}$ is n-dimensional compact internal space's manifold, though potentially too diminutive for local/direct observation, remains an essential component and the simplest example for a compact $n$-dimensional space is the $n$-dimensional torus, viz. $\mathcal{K}^{n}=\mathcal{T}^{n}$, that we consider in this study. We define, on this manifold, a spatially homogeneous but not necessarily isotropic $(1+3+n)$-dimensional synchronous spacetime metric that involves $3$-dimensional spatially flat totally anisotropic external space for $\mathcal{M}^3$ and $n$-dimensional spatially flat isotropic internal space $\mathcal{T}^{n}$:
\begin{equation}
\begin{aligned}
\label{eqn:metric}
{\rm d}S^2=&-{\rm d}t^2+a(t)^2{\rm d}x^{2}+b(t)^2{\rm d}y^{2}+c(t)^2 {\rm d}z^{2} \\
&+s(t)^2\left({\rm d}\theta_{1}^{2}+...+ {\rm d}\theta_{n}^{2}\right), \\
\end{aligned}
\end{equation}
 where $a(t)$ is the scale factor along the $x-$axis and $b(t)$ and $c(t)$ are the scale factors along the  $y-$ and $z-$axes of the $3$-dimensional external space, while $s(t)$ is the scale factor of the $n$-dimensional flat internal space. We describe the $(1+3+n)$-dimensional fluid with an EMT that yields different pressures in the external and internal spaces: 
\begin{equation}
\label{eqn:EMT}
{\tilde{T}_{\mu}}^{\nu}={\textnormal{diag}}[-\tilde{\rho}, \tilde{p}_{\rm ext},\tilde{p}_{\rm ext},\tilde{p}_{\rm ext},\tilde{p}_{\rm int},....],
\end{equation}
where ${\tilde{\rho}}$, $\tilde{p}_{\rm ext}$ and $\tilde{p}_{\rm int}$ are energy density and the isotropic pressures associated with the external and internal spaces as a function of cosmic time. 

The $(1+3+n)$-dimensional Einstein field equations given in \eqref{eqn:EFE} in the presence of the co-moving fluid represented by the EMT (\ref{eqn:EMT}) for the space-time described by the metric \eqref{eqn:metric} are obtained as
\begin{align}
\label{eqn:EFE1}
&\frac {\dot{a}}{a}\frac {\dot{b}}{b}+\frac {\dot{a}}{a}\frac {\dot{c}}{c}+\frac{\dot{b}}{b}\frac{\dot{c}}{c}+n\frac{\dot{s}}{s}\left(\frac {{\dot{\it a}}}{a}+\frac{{\dot{\it b}}}{b}+\frac {{\dot{\it c}}}{c}\right)+\frac{1}{2}n(n-1){\frac{{{\dot{\it s}}}^{2}}{{s}^{2}}}\nonumber \\
&=\tilde{\kappa}\tilde{\rho},\\
\label{eqn:EFE2}
&{\frac {{\ddot{\it b}}}{b}}+{\frac {{\ddot{\it c}}}{c}}+\frac {{\dot{\it b}}}{b}\frac {{\dot{\it c}}}{c}+n\,{\frac {{\ddot{\it s}}}{s}
}+n\,\frac{\dot{s}}{s} \left(\frac{\dot{\it b}}{b}
+\,\frac{\dot{\it c}}{c}\right)+\frac{1}{2} n (n-1)\,{\frac {{{\dot{\it s}}}^{2}}{{s}^{2
}}}\nonumber \\
&=-\tilde{\kappa} \tilde{p}_{\rm ext},\\
\label{eqn:EFE3}
&{\frac {{\ddot{\it a}}}{a}}+{\frac {{\ddot{\it c}}}{c}}+\frac {{\dot{\it a}}}{a}\frac {{\dot{\it c}}}{c}+n\,{\frac {{\ddot{\it s}}}{s}
}+n\,\frac{\dot{s}}{s} \left(\frac{\dot{\it a}}{a}
+\,\frac{\dot{\it c}}{c}\right)+\frac{1}{2} n (n-1)\,{\frac {{{\dot{\it s}}}^{2}}{{s}^{2
}}}\nonumber \\
&=-\tilde{\kappa}\tilde{p}_{\rm ext},\\
\label{eqn:EFE4}
&{\frac {{\ddot{\it a}}}{a}}+{\frac {{\ddot{\it b}}}{b}}+\frac {{\dot{\it a}}}{a}\frac {{\dot{\it b}}}{b}+n\,{\frac {{\ddot{\it s}}}{s}
}+n\,\frac{\dot{s}}{s} \left(\frac{\dot{\it a}}{a}
+\,\frac{\dot{\it b}}{b}\right)+\frac{1}{2} n (n-1)\,{\frac {{{\dot{\it s}}}^{2}}{{s}^{2
}}}\nonumber \\
&=-\tilde{\kappa} \tilde{p}_{\rm ext}, 
\end{align}
and 
\begin{align}
&(n-1)\left({\frac {{\ddot{\it s}}}{s}}+{\frac {{\dot{\it s}}\,{\dot{\it a}}}{sa}}+\,{
\frac {{\dot{\it s}}\,{\dot{\it b}}}{sb}}+{\frac {{\dot{\it s}}\,{\dot{\it c}}}{sc}}+\frac{n}{2}{
\frac {{{\dot{\it s}}}^{2}}{{s}^{2}}}
\right)+{\frac {{\ddot{\it a}}}{a}}+{\frac {{\ddot{\it
b}}}{b}}+{\frac {{\ddot{\it c}}}{c}}\nonumber \\
&+{\frac{{\dot{\it b}}\,{\dot{\it a}}}{ab}}+{
\frac {{\dot{\it c}}\,{\dot{\it a}}}{ac}}+{\frac {{\dot{\it c}}\,{\dot{\it b}}}{bc}}=-\tilde{\kappa} \tilde{p}_{\rm int}.
\label{eqn:EFE5}
\end{align}

 The system given in \eqref{eqn:EFE1}-\eqref{eqn:EFE5} has five differential equations and seven unknown functions $a$, $b$, $c$, $s$, $\tilde{\rho}$, $\tilde{p}_{\rm ext}$ and $\tilde{p}_{\rm int}$, therefore two equations are required to solve the system of field equations to fully determine the system.
We use correlation between the external and internal space kinematics as a constraint and we require an equation to close the system. 

The effective energy density and pressures in Eqs.\eqref{eqn:EFE1}-\eqref{eqn:EFE4} are 
\begin{align}
\label{eqn:EFE1d1}
\tilde{\rho}_{\rm eff}=&\,\frac{1}{\tilde{\kappa}}\left[\tilde{\kappa}\tilde{\rho}-\frac{n(n-1)}{2} \,{\frac{{{\dot{\it s}}}^{2}}{{s}^{2}}}-n\frac{\dot{s}}{s}\bigg(\frac{\dot{a}}{a}+\frac{\dot{b}}{b}+\frac{\dot{c}}{c}\bigg)\right], \\
\label{eqn:EFE2d2}
\tilde{p}_{x,{\rm eff}}=&\frac{1}{\tilde{\kappa}}\left[\tilde{\kappa}\tilde{p}_{\rm ext}+n\frac{\ddot{s}}{s}+\frac{n (n-1)}{2}  \frac{\dot{s}^2}{s^2}+n\frac{\dot{s}}{s}\bigg(\frac{\dot{b}}{b}+\frac{\dot{c}}{c}\bigg)\right],\\
\label{eqn:EFE3d3}
\tilde{p}_{y,{\rm eff}}=&\frac{1}{\tilde{\kappa}}\left[\tilde{\kappa}\tilde{ p}_{\rm ext}+n\frac{\ddot{s}}{s}+\frac{n (n-1)}{2}  \frac{\dot{s}^2}{s^2}+n\frac{\dot{s}}{s}\bigg(\frac{\dot{a}}{a}+\frac{\dot{c}}{c}\bigg)\right],\\
\label{eqn:EFE4d4}
\tilde{p}_{z,{\rm eff}}=&\frac{1}{\tilde{\kappa}}\left[\tilde{\kappa} \tilde{p}_{\rm ext}+n\frac{\ddot{s}}{s}+\frac{n (n-1)}{2}\frac{\dot{s}^2}{s^2}+n\frac{\dot{s}}{s}\bigg(\frac{\dot{b}}{b}+\frac{\dot{a}}{a}\bigg)\right].
\end{align}
The latter terms including $\frac{\dot{s}}{s}$ multiplied by $\frac{\dot{a}}{a}$, or $\frac{\dot{b}}{b}$, $\frac{\dot{c}}{c}$ in the effective directional pressure definitions given in \eqref{eqn:EFE2d2}-\eqref{eqn:EFE4d4} are different in each coordinate directions, though the pressure of the fluid is taken isotropic, viz. $\tilde{p}_{x}=\tilde{p}_{y}=\tilde{p}_{z}=\tilde{p}_{\rm ext}$. These terms disappear in the case that the internal space is static, viz. $\dot{s}=0$, Universe turns out to be isotropic. Modified gravity theories can be recast in the standard GR form such that where all the new geometrical terms are grouped (on the r.h.s.) to form an effective DE contribution denoted as $T_{\mu\nu,{\rm DE}}$, see   \cite{Gannouji:2006jm} for a detailed discussion. As it induces non-zero anisotropic stresses that lead to corresponding shear scalar evolutions, see, e.g., Refs.~\cite{Pimentel89,Madsen88,Faraoni:2018qdr,Akarsu:2019pvi,Mimoso:1995ge}. For instance, as discussed in \cite{Akarsu:2019pvi}, geometrical term makes dark energy dynamical and direction dependent, noting that the shear scalar is kept on the left-hand side as it is a part of the Einstein tensor, namely, the metric itself only. Following the same analogy, as in the dynamics of the Jordan scalar, dynamics of internal spaces contribute to effective energy-momentum source in the four dimensional universe. Since the shear scalar associated with the external space is controlled by the total volume element of higher dimensional space, $\sigma^2\propto V_{\rm tot}^{-2}$ where the total volume element $V_{\rm tot}=V_{\rm int}V_{\rm ext}$ is the volume element of the external (observable) space multiplied by the internal space volume element. For instance, if the universe is higher dimensional, provided that the higher dimensional Universe is steady-state \cite{Akarsu:2021fie} (total volume element is constant) shear scalar may not even evolve, $\sigma^2=V_{\rm tot}^{-2}={\rm const.}$  still staying loyal to GR.
\section{kinematical constraint}
\label{sec:kinconst}
The more general correlation function describing kinematics between the external and internal spaces, which can be postulated as follows:
\begin{equation}
\label{eqn:corr}
\left({\frac{{\dot{\it a}}}{a}}+{\frac {{\dot{\it b}}}{b}}+\frac {{\dot{\it c}}}{c}\right){\frac {{\dot{\it s}}}{s}}=\frac{n}{3}\frac{\dot{V}_{\rm ext}}{V_{\rm ext}}{\frac {{\dot{V}_{\rm int}}}{V_{\rm int}}}=f(t),
\end{equation}
where $f(t)=0$ gives static internal space as in Kaluza-Klein reduction \cite{Kaluza:1921tu,Klein:1926tv} and $f(t)>0$ (or $f(t)<0$) implies that the internal space is expanding (or contracting) coordinately with the expanding external space. A case with the constant rate \cite{akarsuisotropic}, $f(t)=\lambda$, where $\lambda$ is a real constant, substituted in \eqref{eqn:corr} leading
\begin{equation}
\label{eqn:constraint}
\left({\frac {{\dot{\it a}}}{a}}+{\frac {{\dot{\it b}}}{b}}+\frac {{\dot{\it c}}}{c}\right){\frac {{\dot{\it s}}}{s}}=\frac{\lambda}{3n},
\end{equation}
valid for $n\neq0$. The explicit exact solutions of the field equations has been studied in \cite{akarsuisotropic} for $\lambda>0$, since $\frac{n\lambda}{3}$ acts like a cosmological constant \cite{Carroll01,peebles84,Peebles:2002gy}, without introducing the positive and constant $\Lambda$ term to action. They construct $\Lambda$CDM-like cosmological model; with the difference that powers of $t$ in the definition of scale factor are not the same; $a(t)=a_1 \sinh^{\frac{1}{3}}{(\sqrt{\lambda}t)}$ in \cite{akarsuisotropic} for $n=3$---in the presence of three dimensional internal space in addition to three dimensional external space that we observe. On the other hand, in standard $\Lambda$CDM cosmology, the scale factor-time relation holds as
$a(t)=\left(\frac{\Omega_{\rm m}}{\Omega_{\Lambda}}\right)^{\frac{1}{3}}\sinh^{\frac{2}{3}}(\frac{3H_0\sqrt{\Omega_{\Lambda}}t}{2})$. Since the model does not provide as comprehensive an explanation of current universe as the $\Lambda$CDM model, in this study, we also examined whether the number of dimensions could affect the exponent of the sine hyperbolic function using \cite{Akarsu:2014dxa}, more general cosmologies, for arbitrary values of $n$, which have been studied in parametric solution form with the help of Lie symmetry properties.

To investigate expansion anisotropy of external space under kinematical constraint, we will first  accordingly proceed with the definitions of the mean scale factor of the external space as $v_{\rm ext}\equiv V_{\rm ext}^{1/3}$ and that of the internal space as $v_{\rm int}\equiv V_{\rm int}^{1/n}$, where the corresponding volume elements of the 3--dimensional external with volume element $V_{\rm ext}=abc$ and $n$--dimensional internal space with volume element $V_{\rm int}=s^n$. These, in turn, its rate of change of mean scale factor gives the average Hubble parameter of the external space as follows; 
\begin{equation}
\label{Hxdef}
H_{\rm ext}\equiv\frac{\dot{v}_{\rm ext}}{v_{\rm ext}}=\frac{1}{3}\left(H_x+H_y+H_z\right), 
\end{equation}
where $H_{x}=\frac{\dot{a}}{a}$, $H_{y}=\frac{\dot{b}}{b}$ and $H_{z}=\frac{\dot{c}}{c}$ are the directional Hubble parameters along the $x$-- or $y$--axis and the $z$--axis, respectively. Shear scalar is defined by
\begin{align}
\label{eq:sigmaext}
\sigma_{{\rm ext}}^2\equiv &\frac{1}{2}\sum_{i=1}^{3}\left(H_{i}-H_{\rm ext}\right)^{2}\\
=&\frac{1}{3}\left(H_x^2+H_y^2+H_z^2-H_xH_y-H_xH_z-H_yH_z\right).
\end{align}

Similarly, Hubble parameter of internal space reads 
\begin{equation}
H_{\rm int}\equiv\frac{\dot{v}_{\rm int}}{v_{\rm int}}=\frac{{\dot{\it s}}}{s},
\end{equation}
utilising them, we rewrite the kinematical constraint \eqref{eqn:constraint} as
\begin{equation}
\begin{aligned}
\label{eqn:constraintH}
H_{\rm ext}H_{\rm int}=\frac{\lambda}{9} \quad \textnormal{or} \quad \frac{\dot{V}_{\rm ext}}{V_{\rm ext}} \frac{\dot{V}_{\rm int}}{V_{\rm int}}=\lambda.
\end{aligned}
\end{equation}
 We will give in the following section the analytical and explicit Bianchi type I solutions for the choice of $n=3$ considering the higher dimensional fluid given in \eqref{eqn:EMT} as ${\tilde{T}_{\mu}}^{\nu}={\textnormal{diag}}[-\tilde{\rho}, \tilde{p}_{\rm ext},\tilde{p}_{\rm ext},\tilde{p}_{\rm ext},\tilde{p}_{\rm ext},\tilde{p}_{\rm ext},\tilde{p}_{\rm ext}]$, such that the three pressures along external and internal spaces are considered the same. We close the system with six differential equations with six unknown functions $a$, $b$, $c$, $s$, $\tilde{\rho}$ and $\tilde{p}_{\rm ext}$, then discuss the $\lambda>0$ and $\lambda<0$ which govern expanding/contracting internal space at the same rate, respectively accompanied by expanding external space as observations dictate.

\subsection{Positive correlation of spaces ($\lambda>0$)}
\label{sec:positivelambda}
In accordance with observations, external space is expanding, $\lambda>0$ dictates that internal space also expands as like external space as well, explicit solutions of the directional scale factors are as follows:
\begin{align}
\label{eq:scalesols}
a(t)=&a_1\left(\coth{\sqrt{\lambda}t}\right)^{{\frac{k_1}{2\sqrt{\lambda}}}}\sinh^{1/6}{2\sqrt{\lambda}t},\\
b(t)=&b_1\left(\coth{\sqrt{\lambda}t}\right)^{{\frac{k_2}{2\sqrt{\lambda}}}}\sinh^{1/6}{2\sqrt{\lambda}t}, \\
c(t)=&c_1 \left(\coth{\sqrt{\lambda}t}\right)^{{-\frac{k_1+k_2}{2\sqrt{\lambda}}}}\tanh^{1/2}{\sqrt{\lambda}t}\,\, \sinh^{1/6}{2\sqrt{\lambda}t},
\end{align}
where $a_1$, $b_1$, $c_1$, $k_1$ and $k_2$  are positive integration constants satisfying $v_{\rm ext1}=(a_1b_1c_1)^{1/3}$. This solution has the same mathematical form with the four-dimensional anisotropic generalization of the de Sitter Universe, studied in \cite{Grön85}  and was obtained from the asymp- totic form of the Kasner solution for large $t$ values in the presence of the cosmological constant.

Accordingly, the scale factor of internal space reads
\begin{equation}
\label{eq:sdef}
s(t)=s_1\cosh^{1/3}{\sqrt{\lambda}t},
\end{equation}
where $s_1$ is a positive integration constant. Solution of energy density ($\tilde{\rho}$) and the mutual pressure associated with both the external and internal spaces $\tilde{p}_{\rm ext}$ are, respectively, as follows:
\begin{equation}
\begin{aligned}
\label{eq:rhohigher}
\tilde{\rho}=&\frac{1}{\tilde{\kappa}}\left(\frac{5\lambda}{3}+\frac{4\lambda}{3}\frac{1}{\sinh^2{(2\sqrt{\lambda}t)}}-\sigma_{\rm ext}^2\right), 
\end{aligned}
\end{equation}
\begin{equation}
\begin{aligned}
\label{eq:phigher}
\tilde{p}_{\rm ext}=&\frac{1}{\tilde{\kappa}}\left(-\frac{5\lambda}{3}+\frac{4\lambda}{3}\frac{1}{\sinh^2{(2\sqrt{\lambda}t)}}-\sigma_{\rm ext}^2\right),
\end{aligned}
\end{equation}
where 
\begin{align}
\label{eq:sigmaext}
\sigma_{{\rm ext}}^2=\frac{\lambda+3(k_1+k_2)\sqrt{\lambda}+3(k_1^2+k_2^2+k_1k_2)}{3\sinh^2{(2\sqrt{\lambda}t)}}.
\end{align} 
The equation of state (EoS) parameter of the higher dimensional fluid is 
\begin{equation}
\begin{aligned}
\label{eq:eos}
\tilde{w}_{\rm ext}=\frac{-5\lambda+4\lambda{\rm cosech}^2(2\sqrt{\lambda}t)-3\sigma_{\rm ext}^2}{5\lambda+4\lambda{\rm cosech}^2(2\sqrt{\lambda}t)-3\sigma_{\rm ext}^2}
\end{aligned}
\end{equation}
having 
\begin{equation}
\begin{aligned}
\label{eq:eostlimit}
\tilde{w}_{\rm ext}\rightarrow&-1 \quad \textnormal{for sufficiently large values of } t, \\
\tilde{w}_{\rm ext}=&1 \quad \quad \textnormal{at} \quad t=0. 
\end{aligned}
\end{equation}

Here, for the derivation of \eqref{eq:sigmaext}, we have used the directional Hubble parameters as follows:
\begin{equation}
\begin{aligned}
\label{eq:hubble}
H_{x}=\frac{\dot{a}}{a}=&\frac{\sqrt{\lambda}}{3}\,\coth{2\sqrt{\lambda} t}- \,\frac{k_1 }{\sinh{\sqrt{\lambda} t}},\\
H_{y}=\frac{\dot{b}}{b}=&\frac{\sqrt{\lambda}}{3}\,\coth{2\sqrt{\lambda} t}-\,\frac{k_2 }{\sinh{2\sqrt{\lambda} t}},\\
H_{z}=\frac{\dot{c}}{c}=&\frac{\sqrt{\lambda}}{3}\,\coth{2\sqrt{\lambda} t}+\frac{k_1+k_2+\sqrt{\lambda}}{\sinh{\sqrt{\lambda} t}},
\end{aligned}
\end{equation}
satisfying \eqref{Hxdef} with Hubble parameters of external and internal spaces as 
\begin{align}
H_{\rm ext}=&\frac{\sqrt {\lambda}}{3}\coth{\sqrt {\lambda}\;t} \quad \textnormal{and} \quad H_{\rm int}=&\frac{\sqrt {\lambda}}{3}\tanh{\sqrt {\lambda}\;t}.
\label{eq:intH}
\end{align}
At this point, to compare the corresponding density parameter of expansion anisotropy, we prefer writing its definition as a function of mean scale factor in terms of the mean scale factor  using $V_{\rm tot}$. Doing so, volume elements of external and internal spaces are 
\begin{equation}
\begin{aligned}
V_{\rm ext}=v_{\rm ext1}^3\sinh{\sqrt{\lambda}t} \quad\textnormal{and} \quad V_{\rm int}=s_1^3\cosh{\sqrt{\lambda}t},
\end{aligned}
\end{equation}
giving $(1+3+3)$-dimensional universe with the volume element,
\begin{equation}
\begin{aligned}
V_{\rm tot}=V_{\rm ext}V_{\rm int}=\frac{V_{\rm tot1}}{2}\sinh{(2\sqrt{\lambda}t)}.
\label{eq:Vall}
\end{aligned}
\end{equation}
where $V_{\rm tot1}=v_{\rm ext1}^3s_1^3$. From external volume element, we write the mean scale factor 
\begin{equation}
\begin{aligned}
\label{eq:vext}
v_{\rm ext}=v_{\rm ext1}\sinh^{1/3}{\sqrt{\lambda}t},
\end{aligned}
\end{equation}
and the total volume element \eqref{eq:Vall} can be written in terms of $v_{\rm ext}$ as 
\begin{equation}
\begin{aligned}
V_{\rm tot}= V_{\rm tot1}\sqrt{1+\left(\frac{v_{\rm ext}}{v_{\rm ext1}}\right)^6} \left(\frac{v_{\rm ext}}{v_{\rm ext1}}\right)^3.
\label{eq:totV}
\end{aligned}
\end{equation}
Then, subtracting Eq.~(\ref{eqn:EFE2}) from Eq.~(\ref{eqn:EFE3}), and after some manipulations and integration, it turns out that the shear scalar is governed by 
\begin{equation}
\begin{aligned}
\label{eqn:shearprop}
\dot{\sigma}_{\rm ext}+\left(3H_{\rm ext}+nH_{\rm int}\right)\sigma_{\rm ext}=0,
\end{aligned}
\end{equation}
in terms of volume element giving
\begin{align}
\label{eqn:Vtot1}
\frac{\dot{\sigma}_{\rm ext}}{\sigma_{\rm ext}}+\frac{\dot{V}_{\rm tot}}{V_{\rm tot}}=0,
\end{align}
For easier comparison, we prefer to give its definition as a function of mean scale factor in terms of the mean scale factor  using \eqref{eq:totV} in \eqref{eqn:Vtot1} is as follows:
\begin{equation}
\begin{aligned}
\sigma_{\rm ext}^2=V_{\rm tot1}^{-2}\left[1+\left(\frac{v_{\rm ext}}{v_{\rm ext1}}\right)^6\right]^{-1} \left(\frac{v_{\rm ext}}{v_{\rm ext1}}\right)^{-6},
\label{eq:sigmapos}
\end{aligned}
\end{equation}
the alternative written form of \eqref{eq:sigmaext}. When compared to $s=0$ anisotropy, in $(1+3)$-dimensional GR, given as 
\begin{equation}
\begin{aligned}
\sigma_{1+3}^2&\propto v_{1+3}^{-6} \quad \textnormal{for} \quad (1+3)\textnormal{d GR} \\
\sigma_{\rm ext}^2&\propto v_{\rm ext}^{-12} \quad \textnormal{for} \quad (1+3+3)\textnormal{d GR with}\,\, H_{\rm int}H_{\rm ext}=\frac{\lambda}{9},
\end{aligned}
\end{equation}
we have never seen it decay this rapidly here. Although we were accustomed to modifications on shear scalar in modified theories, see \cite{Akarsu:2019pwn,Akarsu:2021max} for Big-bang nucleosynthesis (BBN) and CMB constraints on shear scalar for $\Lambda$CDM model based on GR and \cite{Akarsu:2019pvi} for Brans-Dicke gravity extension of the standard $\Lambda$CDM model, as the shear scalar depends on parameters constrained by local tests in Brans-Dicke type models. This fastest anisotropy decay with the scale factor significantly change the story. It means anisotropies become negligible much earlier in the universe's evolution compared to the standard one, with more stringent constraint on, this aligns the universe closer to isotropy more rapidly, making it appear more like the homogeneous and isotropic FRW model.

If this faster decay occurs during or before recombination, it would lead to a universe that is more isotropic by the time the CMB is emitted. This would reduce any detectable signatures of anisotropy in the CMB, such as quadrupole anisotropies or other non-FLRW effects. Therefore the constraints become stringent. If anisotropies decay much faster, the impact on structure formation may be limited because the universe would transition to isotropy earlier. This could simplify modeling of large-scale structure evolution since deviations from isotropy would be less significant. The shear scalar enters the Einstein field equations through its contribution to the anisotropy of the expansion. If it decays as $v_{\rm ext}^{-12}$, its energy density contribution decreases faster, making it dynamically irrelevant sooner than in the  $v_{\rm ext}^{-6}$ case. This would favor isotropic dynamics at an earlier stage of the universe's evolution.

This may be considered an higher dimensional extension of Wald’s cosmic no-hair theorem \cite{Wald} with positively correlated internal dimensions, and it is very promising for more efficient isotropization. Cosmic no-hair theorem states that Bianchi-type models---except Bianchi IX--- with the EMT of the form $\tilde{\kappa}\tilde{T}_{{\mu\nu}\rm M}+\tilde{\kappa}\tilde{T}_{{\mu\nu},\rm DE},$ where $\tilde{T}_{{\mu\nu},\rm DE}=-\Lambda g_{\mu\nu}+\tilde{T}_{\mu\nu}$, if $\Lambda$ is positive and $\tilde{T}_{\mu\nu}$ satisfies strong and dominant energy conditions, approach de Sitter space exponentially fast, within a few Hubble times $H^{-1}=\sqrt{3/\Lambda}$. CMB measurements indicate that the early Universe have in fact gone through such an accelerated expansion inflationary period. As a result, if cosmic no-hair conjecture/theorem holds, all traces of initial anisotropy should be washed away \cite{Mukhanov,Lyth,Weinberg} and in this model, isotropization is more efficient than the four dimensional GR. 

Let us now the effects of higher dimensional fluid having EoS parameter as \eqref{eq:eos} on the effective four dimensional universe discussing some relaxation occasions to cosmological tensions. Eq. \eqref{eq:eostlimit} reminds us stiff like DE dilutes faster than radiation, in this sense it mimics early dark energy models  \cite{Kamionkowski:2022pkx,Poulin:2018cxd} based on ultra-light axions. Thus, in practice, this model or its extensions may resolve the Hubble tension in a very similar way and effects of these EDE emulators on the CMB spectrum should be investigated comparing with an original EDE parametrizations along with $\Lambda$CDM. As it is known that higher dimensions (as in the modified gravity theories) can be recast in the standard GR form such that $G_{\mu\nu}=\tilde{\kappa}\tilde{T}_{{\mu\nu}\rm M}+\tilde{\kappa}\tilde{T}_{{\mu\nu},\rm DE}$ where all the new geometrical terms are grouped (on the r.h.s.) to form an effective DE contribution (see \cite{Gannouji:2006jm} for details). We note that the shear scalar is kept on the left hand side as it is a part of the Einstein tensor and dynamics of the internal space contribute as DE with anisotropic pressures. Accordingly, we define the DE from Eqs.~\eqref{eqn:EFE1}-\eqref{eqn:EFE4} in the following way :
\begin{align}
3H_{\rm ext}^2-\sigma_{\rm ext}^2=&\tilde{\kappa}\tilde{\rho}+\tilde{\kappa}\tilde{\rho}_{\rm DE},\\
-2\dot{H}_{\rm ext}-3H_{\rm ext}^2-\sigma_{\rm ext}^2=&\tilde{\kappa}\tilde{p}_{\rm ext}+\tilde{\kappa}\tilde{p}_{{\rm DE},x},\\
-2\dot{H}_{\rm ext}-3H_{\rm ext}^2-\sigma_{\rm ext}^2=&\tilde{\kappa}\tilde{p}_{\rm ext}+\tilde{\kappa}\tilde{p}_{{\rm DE},y},\\
-2\dot{H}_{\rm ext}-3H_{\rm ext}^2-\sigma_{\rm ext}^2=&\tilde{\kappa}\tilde{p}_{\rm ext}+\tilde{\kappa}\tilde{p}_{{\rm DE},z},
\end{align}
where $\tilde{\rho}_{\rm DE}$ is the energy density and $\tilde{p}_{{\rm DE},x} $, $\tilde{p}_{{\rm DE},y}$ and $\tilde{p}_{{\rm DE},z}$ are the directional pressures of DE along the $x$- axis and the $y$- and $z$-axes, respectively, which reads
\begin{align}
\label{eqn:2EFErhode}
\tilde{\rho}_{\rm DE}=&\frac{1}{\tilde{\kappa}}\left(-\frac{4\lambda}{3}+\frac{\lambda}{3}\frac{1}{\cosh^2{(\sqrt{\lambda}t)}}\right),\\
\label{eqn:2EFEp1de}
\tilde{p}_{{\rm DE},x}=&\frac{1}{\tilde{\kappa}}\left(\frac{4\lambda}{3}+\frac{\sqrt{\lambda}k_1+\lambda}{2\cosh^2{(\sqrt{\lambda}t)}}\right), \\
\label{eqn:EFE5de}
\tilde{p}_{{\rm DE},y}=&\frac{1}{\tilde{\kappa}}\left(\frac{4\lambda}{3}+\frac{\sqrt{\lambda}k_2+\lambda}{2\cosh^2{(\sqrt{\lambda}t)}}\right), \\
\label{eqn:EFEzde}
\tilde{p}_{{\rm DE},z}=&\frac{1}{\tilde{\kappa}}\left(\frac{4\lambda}{3}+\frac{\lambda-\sqrt{\lambda}(k_1+k_2)}{2\cosh^2{(\sqrt{\lambda}t)}}\right).
\end{align}
Accordingly, the corresponding directional EoS parameters of dark energy, $\tilde{w}_{{\rm DE},i}=\tilde{p}_{{\rm DE},i}/\tilde{\rho}_{{\rm DE},i}$,
\begin{align}
\label{eqn:wdex}
\tilde{w}_{{\rm DE},x}=& 
\frac{\frac{4\lambda}{3}+\frac{\lambda+k_1\sqrt{\lambda}}{2\cosh^2{(\sqrt{\lambda}t)}}}{-\frac{4\lambda}{3}+\frac{\lambda}{3}\frac{1}{\cosh^2{(\sqrt{\lambda}t)}}},\\
\label{eqn:wdey}
\tilde{w}_{{\rm DE},y}=&
\frac{\frac{4\lambda}{3}+\frac{\lambda+k_2\sqrt{\lambda}}{2\cosh^2{(\sqrt{\lambda}t)}}}{-\frac{4\lambda}{3}-\frac{\lambda}{3}\frac{1}{\cosh^2{(\sqrt{\lambda}t)}}},\\
\label{eqn:wdez}
\tilde{w}_{{\rm DE},z}=&\frac{\frac{4\lambda}{3}+\frac{\lambda-(k_1+k_2)\sqrt{\lambda}}{2\cosh^2{(\sqrt{\lambda}t)}}}{-\frac{4\lambda}{3}+\frac{\lambda}{3}\frac{1}{\cosh^2{(\sqrt{\lambda}t)}}},
\end{align}

which constitute an anisotropic dark energy source, at sufficiently large values of $t$, isotropizes and acts as the exact cosmological constant, 
\begin{align}
\tilde{w}_{{\rm DE},x}=\tilde{w}_{{\rm DE},y}=\tilde{w}_{{\rm DE},z}=&-1 \quad \textnormal{as} \quad  t\rightarrow \infty,
\end{align}
which is independent of the value of $\lambda$ provided that $\lambda>0$.
From \eqref{eqn:wdex}-\eqref{eqn:wdez}, it is evident that at early times $(t \sim 0)$, directional equation of state parameters for DE arising from internal dynamics appear as
\begin{align}
\tilde{w}_{{\rm DE},x}(t\sim 0)=&-\frac{11}{6}+\frac{k_1}{2\sqrt{\lambda}}, \\
\tilde{w}_{{\rm DE},y}(t\sim 0)=&-\frac{11}{6}+\frac{k_2}{2\sqrt{\lambda}}, \\
\tilde{w}_{{\rm DE},z}(t\sim 0)=&-\frac{11}{6}-\frac{k_1+k_2}{2\sqrt{\lambda}},
\end{align}
which show that higher dimensions can act effectively as anisotropic phantom-type fluid (provided that $0<k_1<7\sqrt{\lambda}/3$ and $0<k_2<7\sqrt{\lambda}/3$ satisfying $k_1+k_2>7\sqrt{\lambda}/3$), then we give the anisotropic solution reminding that the fact that the stringent constraints on $k_1$ and $k_2$ due to its aggressive evolution (viz. $v_{\rm ext}^{-12}$) can still within the allowed region satisfying $0<k_1<7\sqrt{\lambda}/3$, $0<k_2<7\sqrt{\lambda}/3$ and $k_1+k_2>7\sqrt{\lambda}/3$.

For $k_1=k_2=0$, we return to the isotropic case that was considered in \cite{akarsuisotropic} for $n=3$, in this study, the effective fluid's quantities are given in \eqref{eqn:EFE1d1}-\eqref{eqn:EFE4d4}, which consist of higher dimensional quantities (given in \eqref{eq:rhohigher} and \eqref{eq:phigher}) and DE definitions (arisen from the internal space dynamics) (given in \eqref{eqn:2EFErhode} and  \eqref{eqn:2EFEp1de}), therefore the effective source has a EoS parameter as
\begin{align}
\tilde{w}_{\rm eff}=\frac{1-\sinh^2(\sqrt{\lambda}t)}{1+\sinh^2(\sqrt{\lambda}t)},
\end{align}
four dimensional effective fluid becomes a conventional cosmological constant at late times, and acts like a stiff-fluid at early times which dilutes faster than radiation, in this sense it may mimic early dark energy models, then tailoring higher dimensional theories may be promising in alleviating the Hubble tension. The solution was given in \cite{akarsuisotropic,Akarsu:2014dxa} and at that time we do not aware of the Hubble tension and early dark energy models has not been proposed. 

\subsection{Negative correlation of spaces   ($\lambda<0$) }
\label{sec:negsol}
\subsubsection{de Sitter Bianchi expansion}
The higher-dimensional Robertson-Walker cosmologies that yield constant higher-dimensional volume element has been studied under the two constraints: (i) the total volume element of the universe is constant, and (ii) the effective energy density is constant. It was studied in \cite{Akarsu:2012am} and has been reviewed and extended in \cite{Akarsu:2021fie} considering the constant higher-dimensional volume element ansatz allowing anisotropic metric instead of isotropic one
\begin{equation}
\label{eq:volrel}
V_{\rm tot}=V_{\rm ext}V_{\rm int}=abcs^n={\rm const.},
\end{equation}
as done in Refs.~\cite{Akarsu:2012am,Akarsu:2012vv,Akarsu:2015csa}, with novel discussions. This seems appealing that higher-dimensional steady state (constant total volume element) universe ansatz \eqref{eq:volrel}, if expansion anisotropy is allowed, leads to a constant shear scalar. In what follows, we will rather show that kinematical constraint, if $\lambda<0$ is chosen, corresponds to this ansatz. Simultaneous contraction of the internal space with expansion of the external space takes place for $\lambda<0$. Among a class of solutions includes complex-valued sinusoidal scale factors; we choose the de Sitter solution, which has an internal space scale factor
 \begin{equation}
\begin{aligned}
\label{eq:hubble}
s(t)= e^{H_{\rm int}t}, 
\end{aligned}
\end{equation}
and the directional scale factors are
\begin{equation}
\begin{aligned}
a(t)=&\,a_0 e^{H_x t}\quad , \quad b(t)=b_0 e^{H_{y} t}\quad,\quad c(t)=c_0 e^{H_z t},
\end{aligned}
\end{equation}
where $H_x$ and $H_y$ are integration constants and $H_z$ depends on $\lambda$, $H_x$ and $H_y$ with a relation 
\begin{equation}
\begin{aligned}
H_{z\pm}=\pm \sqrt{|\lambda|}-H_x-H_y,
\end{aligned}
\end{equation}
 giving expanding (contracting) external space with Hubble parameter as 
\begin{equation}
\begin{aligned}
\label{eq:hubble}
H_{{\rm ext}\pm}=-H_{\rm int}=\pm \frac{\sqrt{|\lambda|}}{3}.
\end{aligned}
\end{equation}
Volume elements of external and internal spaces are
\begin{equation}
\begin{aligned}
V_{\rm ext}=V_{\rm ext0} e^{\sqrt{|\lambda|}t}\quad ,\quad
V_{\rm int}=V_{\rm int 0}e^{-\sqrt{|\lambda|}t},
\end{aligned}
\end{equation}
giving higher dimensional steady-state universe
\begin{equation}
\begin{aligned}
V_{\rm tot}= V_{\rm ext0}V_{\rm int0}={\rm const.}.
\label{eq:Vallsteady}
\end{aligned}
\end{equation}
Substituting \eqref{eq:Vallsteady} in Eq. \eqref{eqn:Vtot1}  in turn implies
\begin{equation}
\begin{aligned}
\label{eqn:shearprop}
\dot{\sigma}_{\rm ext}=0,
\end{aligned}
\end{equation}
for $n=3$. We have given two different characteristics for the expansion anisotropy for different signature of correlation. It can be constant or two times aggressive in its decay along with internal space.
Shear scalar solutions are different constants as follows: 
\begin{equation}
\begin{aligned}
\sigma_{{\rm ext}+}^2=&\frac{|\lambda|}{3}-\sqrt{|\lambda|}(H_x+H_y)-H_xH_y,
\label{eq:avehubble}
\end{aligned}
\end{equation}
and
\begin{equation}
\begin{aligned}
\sigma_{{\rm ext}-}^2
=&\frac{|\lambda|}{3}+(1+\sqrt{|\lambda|})H_xH_y+H_x^2+H_y^2,
\end{aligned}
\end{equation}
for the two solutions. The models that exhibit de Sitter volumetric expansion due to the constant effective energy density. If in \cite{akarsuisotropic}, authors had been considered negative correlation ($\lambda<0$), they would achieve higher dimensional steady-state solutions, on the other hand since it was considered for isotropic case, constant nature of shear scalar would not been still revealed. In \cite{Akarsu:2021fie}, authors have considered the constant higher dimensional volume element ansatz, then they have achieved that the higher-dimensional negative cosmological constant plays the role of the four-dimensional positive cosmological constant and they have showcased that the expansion anisotropy, viz., the shear scalar, of the external space imitates a negative cosmological constant viz. a stiff fluid mimicker when allowed on top of the standard $\Lambda$CDM model. Here we show that kinematical constraint would give higher dimensional steady-state Universe with expansion anisotropy standing still as well. Some features of the Bianchi type-I universes in the presence of a fluid that yields an anisotropic equation of state (EoS) parameter have been discussed in \cite{Akarsu:2010zm} in the context of GR. However, in the current study, we consider higher dimensions and first show that higher dimensions can act effectively as anisotropic fluid, then we give the anisotropic de Sitter expansion solution can be achieved due to higher dimensions.  

\subsubsection{Cycloidal solution - from Big Bang to Big Crunch }
For negative values of $\lambda$, there is another solution which starts with Big Bang at $t=0$ and ends with Big Crunch at $t=\frac{\pi}{\sqrt{|\lambda|}}$. The $t=0$ is not a special instant as the periodic expansion/contraction makes any time not unique. Starting from the solution given in \eqref{eq:sdef}, we can perform the transformation $\lambda \rightarrow -\lambda$, which results in hyperbolic functions being replaced by trigonometric ones, the scale factor of internal space turns out to be 
\begin{equation}
s(t)=s_1\cos^{1/3}{\sqrt{|\lambda}|t},
\end{equation}
where $s_1$ is a positive integration constant and mean scale factor of external space reads
\begin{equation}
\begin{aligned}
\label{eq:sinsol}
v_{\rm ext}=v_{\rm ext1}\sin^{1/3}{\sqrt{|\lambda|}t},
\end{aligned}
\end{equation}
 with Hubble parameters of external and internal spaces as 
\begin{equation}
\begin{aligned}
\label{eq:extH}
H_{\rm ext}=\frac{\sqrt {|\lambda|}}{3}\cot{\sqrt {|\lambda|}\;t}\quad \textnormal{and}\quad H_{\rm int}=-\frac{\sqrt {|\lambda|}}{3}\tan{\sqrt {|\lambda|}\;t},
\end{aligned}
\end{equation}
leading the fact that internal space reaches its maximum size when the external space volume element is zero, then while internal space contracts the external space expands. At some instant $t$, they reach the same size then external space reaches its maximum size, begins contracting and finishes its evolution with the Big Crunch. The directional Hubble parameters are as follows:
\begin{equation}
\begin{aligned}
\label{eq:hubble}
H_{x}=&\frac{\sqrt{|\lambda|} }{3}\,\cot{2\sqrt{|\lambda|} t}+k_3 \,\frac{1}{\sin{2\sqrt{|\lambda|} t}},\\
H_{y}=&\frac{\sqrt{|\lambda|} }{3}\,\cot{2\sqrt{|\lambda|}t}+k_4 \,\frac{1}{\sin{2\sqrt{|\lambda|}  t}},\\
H_{z}=&\frac{\sqrt{|\lambda|}}{3}\,\cot{2\sqrt{|\lambda|} t}-\frac{k_3+k_4+\sqrt{\lambda}}{\sin{2\sqrt{|\lambda|}t}},
\end{aligned}
\end{equation}
here $k_3=ik_1$ and $k_4=ik_2$.
Volume elements of external and internal spaces are
\begin{equation}
\begin{aligned}
V_{\rm ext}=v_{\rm ext1}^3\sin{\sqrt{|\lambda|}t} \quad\textnormal{and} \quad V_{\rm int}=s_1^3\cos{\sqrt{|\lambda|}t},
\end{aligned}
\end{equation}
giving $(1+3+3)$-dimensional universe with the volume element,
\begin{equation}
\begin{aligned}
V_{\rm tot}=V_{\rm ext}V_{\rm int}=\frac{V_{\rm tot1}}{2}\sin{(2\sqrt{|\lambda|}t)},
\label{eq:Vall2}
\end{aligned}
\end{equation}
where $V_{\rm tot1}=v_{\rm ext1}^3s_1^3$, written in terms of $v_{\rm ext}$ as 
\begin{equation}
\begin{aligned}
V_{\rm tot}= V_{\rm tot1}\sqrt{1-\left(\frac{v_{\rm ext}}{v_{\rm ext1}}\right)^6} \left(\frac{v_{\rm ext}}{v_{\rm ext1}}\right)^3,
\label{totV2}
\end{aligned}
\end{equation}
which shows that external space volume element can pass from zero, on the other hand it is not possible for $\lambda>0$, see \eqref{eq:totV} for comparison. 
Substituting \eqref{totV2} in \eqref{eqn:Vtot1}, we obtain 
\begin{equation}
\begin{aligned}
\sigma_{\rm ext}^2=V_{\rm tot1}^{-2}\left[1-\left(\frac{v_{\rm ext}}{v_{\rm ext1}}\right)^6\right]^{-1} \left(\frac{v_{\rm ext}}{v_{\rm ext1}}\right)^{-6}.
\label{avehubble}
\end{aligned}
\end{equation}
The evolution is cycloidal — the scale factor grows at an ever-decreasing rate until it reaches a point at which the expansion is halted and reversed (where the shear scalar is minimum). The Universe then starts to compress and it finally collapses in the Big Crunch for matter/radiation dominated closed Friedman Universe. Friedmann described the oscillation of the Universe as a single cycle from big bang to big crunch for a particular case as well, where the cosmological constant is negative or zero, $\Lambda\leq0$. This solution is similar to Friedmann solution for closed Universe.

\section{Wormhole-type Eucledian continuation $[\lambda \rightarrow -\lambda \,\, \cup \,\, t\rightarrow -i\tau]$}

As a further step, if $\lambda$ changes its signature together with Wick rotation, we see that $\sqrt{\lambda}t$---the argument of hyperbolic function---remains invariant, still satisfying \eqref{eqn:constraintH}. This property is independent of the isotropy, for simplicity
 we will show this continuation in Robertson-Walker metric,  the $(1+3+3)$-dimensional synchronous spacetime metric is given in \eqref{eqn:metric} as
\begin{equation}
\begin{aligned}
\label{eqn:metriccontinuation}
{\rm d}S^2=-{\rm d}t^2+a(t)^2\,{\rm d}\vec{x}^{2}+s(t)^2\,\sum_{n=1}^{3}{\rm d}\vec{\theta}^{\,2}, \\
\end{aligned}
\end{equation}
which has a Lorentzian signature.
The Euclidean solution, which is time-symmetric, arises via analytic continuation with transformation $t\rightarrow-i\tau$ and $\lambda<0$ as follows;
\begin{equation}
\begin{aligned}
{\rm d}S^2={\rm d}\tau^2+a_{\rm E}(\tau)^2\,{\rm d} \,\vec{x}^{2}+s_{\rm E}(\tau)^2\,\sum_{n=1}^{3}{\rm d}\vec{\theta}^{\,2}, \\
\end{aligned}
\end{equation}
which has a Eucledian signature with relations: $a_{\rm E}(\tau)=a(-i\tau)$, $s_{\rm E}(\tau)=s(-i\tau)$, $H_{\rm ext,E}=H_{\rm ext}(-i\tau)$ and $H_{\rm int,E}=H_{\rm int}(-i\tau)$. 
In one region the geometry will be Lorentzian (with signature $-+++$) and in the other it will be Euclidean $(++++)$ \cite{Dereli:1993pj}. We achieve a wormhole-type Eucledian continuation of the Lorentzian solutions if the internal space exists and the external space are correlated with a constant ratio of their Hubble parameters. In semi-analytical approach, the solutions of the classical Euclidean field equations are instantons, yet do exist only in the presence of  special types of matter, see Refs. \cite{Giddings,Lee,Coule}, i.e., for an imaginary minimally coupled massless scalar field shown in \cite{hawkingpage}, not in pure gravity (gravity without being coupled to any other form of matter). In \cite{zhuk93}, it was given some important case examples of the closed RW universe has a wormhole-type Eucledian continuation with wormhole geometry. In \cite{bleyer94}, there is a scalar field minimally coupled (as a degree of freedom to close the system) having a specific potential. A higher dimension or scalar field, provided it is dynamical and follows a scenario in which the internal space first contract and then expand at the same rate, contributes to the field equations in an indistinguishable manner. The constancy of this contribution mimics a cosmological constant, and its sign transition mimics a sign switch of the cosmological constant, allowing for an analytical continuation from the Lorentzian metric to the Euclidean one.

The former negative correlation between the spaces, along with the accelerated expansion of the external space, leads to the contraction of the internal space (solution given in Sec. \ref{sec:negsol}). A possible mechanism, such as a wormhole-type continuation, may reverse the correlation from negative to positive ($\lambda < 0 \rightarrow \lambda > 0$), causing both spaces to begin expanding. This solution, given in Sec.  \ref{sec:positivelambda}, initially follows a sine function [Eq.\eqref{eq:sinsol}] and then evolves into a hyperbolic sine function [Eq.\eqref{eq:vext}] with the same correlation constant $\lambda$. Although we have primarily studied the case of constant $\lambda$, sigmoid functions such as $\tanh(z)$ may better represent this scenario, as they have a non-negative derivative at each point and are constrained by two horizontal asymptotes—for example, almost horizontal for $\lambda$ today and $-\lambda$ after a certain redshift. We have shown that both positive $\lambda$ and negative $\lambda$ solutions can exist.

As discussed in the concluding paragraphs of Sections \ref{sec:kinconst} and \ref{sec:positivelambda}, or as explicitly shown by Eq. \eqref{eq:vext}, the mean scale factor for the external space is given by $v_{\rm ext} = v_{\rm ext1} \sinh^{1/3}{\sqrt{\lambda}t}$, which evolves as $t^{1/3}$ for small $t$ values. This model does not recover the early universe dynamics described by the standard model, where $a \propto t^{2/3}$. This discrepancy prompted us to consider the following question: If the dimension of internal space determines the time dependence of the external space's scale factor, what would happen if we changed the number of dimension of internal space? However, for an arbitrary number of dimension of the internal space, the general solution can only be analyzed in parametric form using the Lie symmetry properties discussed in \cite{Akarsu:2014dxa}, in which the author of this study also contributed. It can be seen from Figures 1(d)-2(d) of \cite{Akarsu:2014dxa} that the effective equation of state parameter $\tilde{w}$ always starts from unity and evolves to $-1$ for $n = 1 \ldots 9$, leading us to conclude that no further progress can be made through this approach.

\section{Conclusion}

In this study, we have extended the work of \cite{akarsuisotropic,Akarsu:2014dxa} allowing anisotropic expansion in the external space and possible contraction of the internal space. In what follows, we have explored in detail the expansion anisotropy and the positive/negative correlation extensions of higher-dimensional Einstein field equations in standard GR, separately or simultaneously, toward theoretically achievable pathways for developing observationally consistent models within a robust theoretical framework. We have given exact solutions to the higher dimensional Einstein field equations considering a spatially homogeneous and flat spacetime, incorporating a three dimensional anisotropic external space and a $n$-dimensional isotropic internal space, under the assumption that the term arises as the multiplication of the Hubble parameters of the internal and external spaces in the modified Friedmann equations remains constant, $H_{\rm int}H_{\rm ext}=\frac{\lambda}{9}$ where $\lambda>0$ and  $\lambda<0$ cases. Incorporating anisotropy into the field equations has enabled us to consider alternative possibilities for even $\lambda>0$, studied in \cite{akarsuisotropic}.

The interest in anisotropic cosmologies has never disappeared, has recently intensified significantly due to recent observations, see \cite{Jones:2023ncn} and references therein. A deviation from the stiff-fluid character of the shear scalar might imply the necessity for replacing $\Lambda$ (or scalar fields) by an anisotropic stress, see Ref.~\cite{Barrow:1997sy} for a list. For positive correlation, if Wald's cosmic no-hair conjecture/theorem holds, all traces of initial anisotropy should be washed away and in this model, isotropization is more efficient than the four dimensional GR. Anisotropy, without taking a role in alleviating tensions, may provide an opportunity to manipulate the CMB quadrupole temperature fluctuation at the desired amount, viz. suggesting a possible answer to the fact that its observed value is lower than that predicted by the standard $\Lambda$CDM model \cite{Bennett11,Ade:2013kta,Schwarz:2015cma,Akrami:2019bkn}. These are the first attempts and our findings signal the capability of higher dimensions in point of constructing models reducing cosmological tensions. 

Alternatively, negative correlation---acting as negative cosmological constant---automatically yields a higher-dimensional steady-state universe leading a constant shear scalar, $\rho_{\sigma^2}=\sigma^2/2={\rm const.}$ in alignment with its definition, $\sigma^2=V_{\rm tot}^{-2}$. Furthermore, some features of the Bianchi type-I universes in the presence of a fluid that yields an anisotropic equation of state parameter have been discussed in \cite{Akarsu:2010zm} in the context of GR, on the other hand, in the current study, we consider higher dimensions and first show that higher dimensions can act effectively as anisotropic fluid, then we give the anisotropic de Sitter expansion solution. For the chosen negative correlation, cycloidal model, which starts with Big Bang and would collapse back into a Big Crunch, is another model arisen from the replacement of the argument of hyperbolic function with the  complex numbers due to $\lambda<0$, e.g., $\cosh{(\sqrt{i^2\lambda})t}=\cosh{(i\sqrt{|\lambda|}t)}=\cos{\sqrt{|\lambda}|t}$. We theoretically explore the signature change of correlation, $\lambda\rightarrow-\lambda$, viz. expanding internal space suddenly becomes contracting, and if Wick rotation (by substitution $t\rightarrow-i\tau$) accompanies this signature change as well, field equations and their solutions remain totally invariant. This showcases an analytical continuation from Lorentzian region into the Eucledian region with a wormhole-type topology associated with $\lambda\rightarrow-\lambda$, broaden our perspective on possible dynamics, including varying correlations modeled by functions of redshift, $f(z)$.  

Another interesting feature is that if the spatial coordinate changes signature; namely the space coordinate (of internal/external spaces) turns into time coordinate, scale factors in front ($a^2$ or $s^2$) remains invariant. Therefore, we could not distinguish this signature change on spatial coordinates in Einstein field equations for flat ($k=0$) space-like sections. Because Hubble parameter and its derivatives---as these contribute with the ratio form to field equations---remains the same leaving Einstein field equations exactly the same, on the other hand,  curvature of space, if exists, contributes to field equations with $\pm 1/a^2$ (or $\pm 1/s^2$) depending on being closed/open spaces. A unique indication of the signature change on spatial coordinates is the spatial curvature, $k_{\rm ext}/a^2$ or $k_{\rm int}/s^2$. In \cite{Alexandre:2023nmh}, classical imaginary space extensions of Lorentzian theory with $a^2<0$ has been studied and cosmological constant switches its sign via a metric signature change across boundaries with a degenerate metric in Einstein-Cartan and Plebanski formulations of general relativity. Similarly in higher dimensional models, we realized that with a sudden change of the scale of the static internal space in the presence of internal curvature ($k_{\rm int}$) is realized as a sudden change of the cosmological constant $\tilde{\Lambda}_{\rm eff}=\tilde{\Lambda}-\frac{n(n-1)k_{\rm int}}{s^2}$---with accompanying abrupt change on the gravitational coupling constant---as observed in the external manifold, a detailed investigation is in progress and will be presented in a future work under progress \cite{curvaturesignswitch}. A flow scheme of the study is given in Fig.~\ref{fig:Scheme} in Appendix \ref{app:scheme} to assist readers in navigating the paper and in relating with the other our studies in literature.
\appendix
\section{The scheme of the study}
\label{app:scheme}
\begin{figure}[ht!]
\begin{center}
\includegraphics[trim =0mm  0mm 0mm 0mm, clip, width=0.50\textwidth]{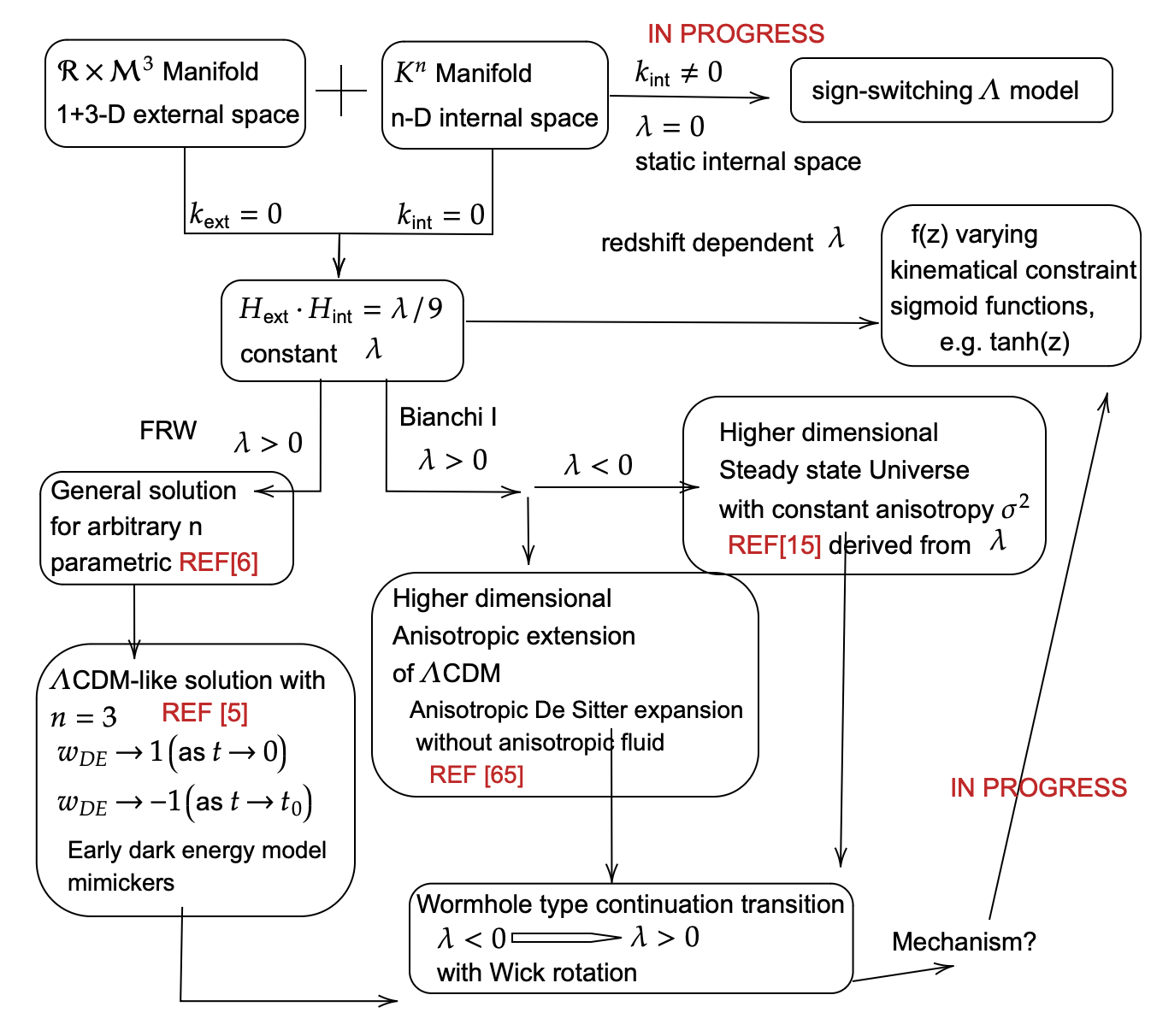}
\end{center}
\caption{The scheme of the study}
\label{fig:Scheme}
\end{figure}

Fig.\ref{fig:Scheme} outlines the structure of our study, ensuring that each research question is addressed accordingly with their relation to previous/future studies. It acts as a blueprint for the research presented in the sections that follow.
\begin{acknowledgments}
The author thanks to \" Ozg\" ur Akarsu for valuable discussion and the anonymous referee for their constructive comments regarding the manuscript. This
study was supported by Scientific and Technological
Research Council of Turkey (TUBITAK) under the Grant Number 122F124. This article is based upon work from COST Action CA21136 Addressing observational tensions in cosmology with systematics and fundamental physics (CosmoVerse) supported by COST (European Cooperation in Science and Technology).
\end{acknowledgments}

\section{Data Availability Statement}
No Data associated in the manuscript. 

\end{document}